\documentclass[a4paper, amsfonts, amssymb, amsmath, reprint, showkeys, nofootinbib, twoside,notitlepage,onecolumn]{revtex4-1}

\usepackage{amsmath,amstext}
\usepackage[T1]{fontenc}
\usepackage{amssymb}
\usepackage{graphicx}
\usepackage{ae,aecompl}

\DeclareFontFamily{OT1}{pzc}{}
\DeclareFontShape{OT1}{pzc}{m}{it}{<-> s * [1.10] pzcmi7t}{}
\DeclareMathAlphabet{\mathpzc}{OT1}{pzc}{m}{it}

\usepackage{hyperref}
\usepackage{amsmath}
\usepackage{amssymb}
\usepackage{mathtools}
\usepackage{bm}
\usepackage{cleveref}
\usepackage{tensor}
\usepackage{braket}
\usepackage{enumitem}
\usepackage{mhchem}
\usepackage{amsthm}
\usepackage{nccmath}
\usepackage{mathrsfs}
\usepackage{color}

\newcommand{\E}{{\mathbb{E}}}
\newcommand{\curlyE}{\mathcal{E}_G}
\newcommand{\tauin}{\tau_\textup{in}}
\newcommand{\tauel}{\tau_\textup{el}}
\newcommand{\tauT}{\tau_{T}}
\newcommand{\feq}{{f_\textup{eq}}}

\newcommand{\PC}{{\mathbb{P}_C}}
\newcommand{\PN}{{\mathbb{P}_N}}

\newcommand{\s}{{\sigma \!\! \! \sigma}}

\def\be{\begin{equation}}
\def\ee{\end{equation}}
\def\beq{\begin{eqnarray}}
\def\eeq{\end{eqnarray}}

\theoremstyle{definition}

\theoremstyle{theorem}

\theoremstyle{corollary}

\begin{document}
\title{Beyond the Onsager matrix: All-order resummation of relativistic particle and energy diffusion}
\author{L.~Gavassino}
\affiliation{Department of Applied Mathematics and Theoretical Physics, University of Cambridge, Wilberforce Road, Cambridge CB3 0WA, United Kingdom}

\begin{abstract}
We study the coupled diffusion of particle number and energy carried by ultrarelativistic probe particles propagating through a material medium. The probes undergo elastic scattering, which randomizes their direction, and inelastic scattering, which exchanges energy with the medium while preserving particle number. We compute exactly the full $2\times2$ frequency- and wave-number-dependent diffusivity matrix, resumming the constitutive gradient expansion to all orders. In a particularly natural Onsager-normalized basis of conserved densities, the resummed matrix is found to be diagonal at arbitrary frequency and wave number, even though particle and energy transport remain coupled and the conventional Onsager matrix has nonzero thermoelectric cross coefficients. Beyond the hydrodynamic limit, the resummed diffusivities resolve microscopic relaxation timescales and the ballistic kinetic continuum that are invisible to the ordinary Onsager coefficients.
\end{abstract} 
\maketitle

\vspace{-0.8cm}
\section{Introduction}
\vspace{-0.3cm}

Diffusion in matter is rarely the transport of a single conserved charge. In many condensed-matter and astrophysical systems, one deals with a rigid material composed of several chemical species, one of which is much more mobile than the others. A local perturbation of the mobile component then carries both particle number and energy. In this setting, momentum is not a relevant hydrodynamic variable because of the assumed rigidity of the medium, while the long-wavelength dynamics generally involves the coupled diffusion of multiple conserved quantities. This is precisely the type of transport problem described by Onsager's theory of irreversible processes \cite{Onsager_1931}. Familiar examples include electronic transport in solids and neutrino transport through dense neutron-star matter \cite{prigoginebook,BurrowsReddyThompson2006,MadsenSingh2006,Prakash2001,GavassinoThermoelectric:2025bnx}. Analogous coupled diffusion problems also arise in relativistic many-body systems, such as the simultaneous transport of baryon number, strangeness, and electric charge in the quark-gluon plasma \cite{Greif2018,Fotakis2021,Das2022}.

Let $\delta n$ denote the perturbation of the particle density of the mobile
species and $\delta\varepsilon$ the perturbation of the \textit{total} energy density.
At sufficiently long wavelengths, their linearized evolution takes the
general form
\begin{equation}
\partial_t
\begin{bmatrix}
\delta n \\
\delta \varepsilon
\end{bmatrix}
=
\begin{bmatrix}
\mathfrak{D}_{nn} & \mathfrak{D}_{n\varepsilon} \\
\mathfrak{D}_{\varepsilon n} & \mathfrak{D}_{\varepsilon\varepsilon}
\end{bmatrix}
\partial_x^2
\begin{bmatrix}
\delta n \\
\delta \varepsilon
\end{bmatrix}+\text{``Sources''}.
\label{eq:coupled_diffusion}
\end{equation}
The matrix $\mathfrak{D}_{IH}$ is related to the usual Onsager matrix by a
thermodynamic change of variables. Its diagonal entries describe particle
and energy diffusion, while its off-diagonal entries encode the coupling
between the two transport processes.
Equation~\eqref{eq:coupled_diffusion}, however, is only the leading-order
description in a derivative expansion. It assumes that the characteristic
wavelength and time scale of the perturbation are large compared with the
microscopic mean free path and relaxation time of the mobile species. At shorter scales, higher
derivatives become important, and the four entries of the diffusion matrix
must themselves be promoted to functions of frequency and wave number \cite{ChungYip1969,Hess1983,TGaskell_1987,VelascoColin,Hansen2007,AlfordBulk,Bu:2014sia,StephanovHydroPlus:2017ghc,Kiselev,BulkGavassino,HellerSingulant:2021yjh,Krihin2023,GavassinoInfiniteOrded2024pgl,Nazaryan2024,GavassinoDiamond:2026fgz},
\begin{equation}
\mathfrak{D}_{IH}
\quad\longrightarrow\quad
\mathfrak{D}_{IH}(i\omega,ik),
\label{eq:resummed_matrix}
\end{equation}
where the usual diffusion matrix is recovered as
$\mathfrak{D}_{IH}(0,0)$, whereas the full functions
$\mathfrak{D}_{IH}(i\omega,ik)$ resum the derivative expansion to all orders.

The question we address in this work is simple: can the full
$2\times2$ conduction matrix be determined exactly, beyond the regime in
which the ordinary Onsager description applies? We answer this question affirmatively for
a solvable relativistic kinetic model consisting of ultrarelativistic probe
particles propagating through a rigid material medium. The particles undergo
both elastic collisions, which randomize their direction of propagation, and
inelastic collisions, which exchange energy with the medium while preserving
particle number. The resulting system has two conserved quantities: the
number of probe particles and the total energy of particles plus the medium.
We compute the complete matrix $\mathfrak{D}_{IH}(i\omega,ik)$ in closed
form. The result provides an all-order description of coupled particle and
energy diffusion, interpolating between the long-wavelength diffusive regime
and kinetic free streaming, and retaining physical information that is invisible to
the ordinary Onsager coefficients. 

Throughout the article, we adopt the metric signature $(-,+,+,+)$, and work in natural units, $c=\hbar=k_B=1$. The system of interest lives in three spatial dimensions, but all gradients are assumed to point in the direction $x$.

\vspace{-0.3cm}
\section{Resummed diffusivities: A quick recap}\label{quickrecap}
\vspace{-0.3cm}

Before turning to the problem at hand, it is useful to briefly review the
abstract formalism that allows one to express the resummed diffusivities
$\mathfrak{D}_{IH}(i\omega,ik)$ directly in terms of microscopic operators.
We will restrict ourselves to the ingredients needed for the calculation
below, referring the reader to \cite{GavassinoDiamond:2026fgz} for a more complete discussion.

Consider a linear system of the form
\begin{equation}
\partial_t \Psi+ \E\partial_x\Psi +\s \Psi=\text{``Sources''}\, ,
\label{eq:abstract_kinetic}
\end{equation}
where $\Psi:\text{``Minkowski''} \to \mathcal{H}$ is a linearized kinetic degree of freedom taking values in a Hilbert space $\mathcal H$, equipped with the Onsager inner product $(\cdot,\cdot)$. The latter is defined by the condition that the local quadratic free-energy density is $\Delta\mathfrak{F}=\frac{1}{2}(\Psi,\Psi)$. The operators $\E$ and $\s$ are both self-adjoint in $\mathcal{H}$ due to Onsager reciprocity \cite{GavassinoSymmetric2022nff,GavassinoDisturbing:2026klp}, with $\E$ the streaming operator and $\s$ a non-negative collision operator.

The Hilbert space naturally decomposes as $\mathcal{H}=\ker(\s)\oplus[\ker(\s)]^\perp\equiv\mathcal{H}_C\oplus\mathcal{H}_N$, with $\mathcal{H}_C$ the space of conserved degrees of freedom and $\mathcal{H}_N$ the space of non-conserved degrees of freedom. Choosing an orthonormal basis $\{e_I\}$ of $\mathcal{H}_C$, we have the orthogonal projectors $\PC=\sum_I e_I (e_I,\cdot)$ and $\PN=1-\PC$ onto the respective subspaces.

Taking the inner product of \eqref{eq:abstract_kinetic} with $e_I$, we obtain the conservation laws
\begin{equation}
\partial_t \rho_I+\partial_x J_I=\text{``Sources''} \, ,
\end{equation}
with conserved densities $\rho_I=(e_I,\Psi)$ and conserved fluxes $J_I=(e_I,\E\Psi)$. If, instead, we apply $\PN$ to both sides of \eqref{eq:abstract_kinetic}, invoke the orthogonal decomposition $\Psi=\PC\Psi+\PN\Psi=\sum_H \rho_H e_H+\Psi_N$, and assume that both $\Psi$ and the sources have a spacetime dependence of the form $e^{ikx-i\omega t}$, we can solve for $\Psi_N$ to obtain
\begin{equation}\label{PsiN}
\Psi_N=-ik\sum_H [\s_N-i\omega+ik\E_N]^{-1} \PN\E e_H \rho_H +\text{``Sources''} \, ,
\end{equation}
where $\s_N=\PN \s \PN$ and $\E_N=\PN\E\PN$ are compressed operators, which should always be regarded as operators on $\mathcal{H}_N$.
Plugging \eqref{PsiN} into the formula for the fluxes $J_I$, and assuming that $(e_I,\E e_H)=0$ (which will be the case in our system of interest), we finally obtain $J_I=-ik\sum_H \mathfrak{D}_{IH}(i\omega,ik)\rho_H+\text{``Sources''}$, with
\begin{equation}\label{DIHGeneral}
\mathfrak{D}_{IH}(i\omega,ik)= (\PN\E e_I,[\s_N-i\omega+ik\E_N]^{-1}\PN \E e_H) \, .
\end{equation}
These are the resummed diffusivities we are interested in. In the limit $\omega,k\rightarrow0$, they reduce to the ordinary diffusion matrix, while their dependence on $\omega$ and $k$ resums the constitutive derivative expansion to all orders.
Finally, since $\s_N$ and $\E_N$ are self-adjoint, we have $\mathfrak{D}_{IH}(i\omega,ik)=\mathfrak{D}_{HI}^*((i\omega)^*,(ik)^*)$, which is an instance of Onsager symmetry.

\vspace{-0.1cm}
\section{Physical setup}
\vspace{-0.1cm}

We now introduce the kinetic model that we shall study in this paper. We will recast its equations in the general form \eqref{eq:abstract_kinetic}, identify the relevant conserved vectors $e_I$, and determine the conservation laws associated with them.

\vspace{-0.1cm}
\subsection{Kinetic equations}
\vspace{-0.1cm}

We consider a rigid material medium at rest, whose constituents have a negligibly short mean free path. The medium is always in local thermodynamic equilibrium within itself, and its constituents are fixed at their locations, so that the only relevant degree of freedom we assign to it is its temperature perturbation $\delta T(t,x)$. Transport is mediated by a gas of ultrarelativistic particles (which may be bosons or fermions), whose mean free path is much longer and whose particle number is conserved. We will track the full perturbation $\delta f(t,x,p^j)$ of the kinetic distribution function of the gas, which counts the occupation of one-particle states at the spacetime point $(t,x)$ with momentum $p^j$.

The equations of motion of the system are postulated to be
\begin{equation}\label{kineticmodel}
\begin{split}
& c_M \partial_t \delta T +\int \dfrac{gd^3p}{(2\pi)^3} p^0 (\partial_t {+}v^x\partial_x) \delta f =0 \, , \\
& (\partial_t {+}v^x\partial_x) \delta f  =\dfrac{1}{\tauin} [\feq(1{+}s\feq)(\delta\alpha+\beta^2 p^0\delta T) -\delta f]+\dfrac{1}{\tauel} (\langle \delta f\rangle -\delta f) \, .
\end{split}
\end{equation}
The first line is simply the conservation law $\partial_\mu \delta T^{\mu 0}=0$ for the total ``medium''+``gas'' system, with $c_M$ the heat capacity per unit volume of the medium and $g$ the spin/color/flavor degeneracy of the gas. The second line is the Boltzmann equation for the gas. Here, $p^\alpha$ is the four-momentum, with $p^\alpha p_\alpha=0$, and $v^j=p^j/p^0\in S^2$. The particles of the gas do not interact with each other, but collide with the medium. The first collision term describes inelastic scattering, which exchanges energy between the gas and the medium, while the second describes elastic scattering, which preserves the energy of each gas particle. The corresponding relaxation timescales are $\tauin$ and $\tauel$, which we assume to be constant. 

The two contributions to the collision operator deserve closer inspection.
The inelastic processes are assumed to be fully randomizing, so that the corresponding collision term takes a relaxation-time-approximation form, i.e.\ $\propto\delta\feq-\delta f$, where $\delta\feq=\feq(1{+}s\feq)(\delta\alpha+\beta^2p^0\delta T)$ is the local Maxwell-Boltzmann/Bose-Einstein/Fermi-Dirac distribution that the collisions tend to establish. Here, $\feq=(e^{-\alpha+\beta p^0}{-}s)^{-1}$, with $s=0,1,-1$ respectively, and $\beta=1/T$. Since collisions are mediated by the medium, this local equilibrium state describes a gas at rest at the local temperature $T+\delta T$ \textit{of the medium}, as occurs, for example, in radiative-transfer problems governed by Kirchhoff-Planck relations \cite{mihalas_book}. However, since the collisions do not change the number of gas particles, $\delta\feq$ and $\delta f$ must have the same particle density. This fixes the local-equilibrium fugacity perturbation through the matching condition
\begin{equation}\label{matchingconditiondeltaalpha}
\delta \alpha =\dfrac{\displaystyle\int d^3 p \, [\delta f-\feq (1{+}s\feq)\beta^2 p^0 \delta T]}{\displaystyle\int d^3p \, \feq (1{+}s\feq)} \, .
\end{equation}

The elastic processes are also assumed to be fully randomizing (equivalently, isotropic). Hence, the corresponding collision term takes the standard form $\langle\delta f\rangle-\delta f$, where $\langle\cdot\rangle$ denotes the angular average at fixed energy. In particular,
\begin{equation}
\langle \delta f\rangle |_{p^0}=\int_{S^2} \dfrac{d^2 \Omega}{4\pi} \delta f(p^0 v^j) \, .
\end{equation}

\subsection{Onsager normalization}

In the present system, the quadratic free-energy perturbation reads \cite{GavassinoCausality2021,RochaGavassinoFlucut:2024afv}
\begin{equation}
2\Delta\mathfrak{F}=\beta c_M (\delta T)^2 +\int \dfrac{gd^3 p}{(2\pi)^3} \dfrac{(\delta f)^2}{\feq (1{+}s\feq)\beta} \, .
\end{equation}
This suggests that we introduce the Onsager-normalized degree of freedom
\begin{equation}
\Psi=
\begin{bmatrix}
\Psi_T \\
\psi \\
\end{bmatrix}
= 
\begin{bmatrix}
\sqrt{\beta c_M}\, \delta T \\
\\
\sqrt{\frac{g}{(2\pi)^3 \feq (1+s\feq)\beta}} \, \delta f \\
\end{bmatrix} \, .
\end{equation}
In this way, the Onsager inner product is simply $(\Phi,\Psi)=\Phi_T^*\Psi_T+\int d^3 p \, \phi^* \psi$, so that $\mathcal{H}=\mathbb{C}\oplus L^2(\mathbb{R}^3)$. For convenience, we will also denote $(\phi,\psi)=\int d^3 p \, \phi^* \psi$, so that $(\Phi,\Psi)=\Phi_T^*\Psi_T+(\phi,\psi)$.

With this choice of variables, and defining
\begin{equation}\label{definechi}
\chi =\sqrt{\frac{g \feq(1{+}s\feq)}{(2\pi)^3\beta}} \, ,
\end{equation}
equation \eqref{kineticmodel} takes the standard form \eqref{eq:abstract_kinetic}, now with no source terms, 
\begin{equation}
\begin{split}
&  \partial_t \Psi_T - \dfrac{1}{\tauin\sqrt{c_M T^3}}  \left[ (p^0\chi,\psi-\tfrac{p^0\chi \Psi_T}{\sqrt{c_M T^3}})- \dfrac{(\chi, \psi- \tfrac{p^0\chi\Psi_T}{\sqrt{c_M T^3}})}{(\chi, \chi)} (p^0\chi,\chi) \right]=0
\, , \\
& (\partial_t {+}v^x\partial_x)\psi  +\dfrac{1}{\tauin} \left[\psi-\tfrac{p^0\chi\Psi_T}{\sqrt{c_M T^3}}- \dfrac{(\chi, \psi- \tfrac{p^0\chi\Psi_T}{\sqrt{c_M T^3}})}{(\chi, \chi)} \chi \right]+\dfrac{1}{\tauel} (\psi-\langle \psi\rangle)=0 \, ,\\
\end{split}
\end{equation}
from which we can immediately read off the relevant operators:
\begin{equation}
\E
\begin{bmatrix}
\Psi_T \\
\psi \\
\end{bmatrix}=
\begin{bmatrix}
0 \\
v^x \psi \\
\end{bmatrix} \, ,
\qquad \qquad \s
\begin{bmatrix}
\Psi_T \\
\psi \\
\end{bmatrix}=
\begin{bmatrix}
- \dfrac{1}{\tauin\sqrt{c_M T^3}}  \left[ (p^0\chi,\psi-\tfrac{p^0\chi\Psi_T}{\sqrt{c_M T^3}})- \dfrac{(\chi, \psi- \tfrac{p^0\chi\Psi_T}{\sqrt{c_M T^3}})}{(\chi, \chi)} (p^0\chi,\chi) \right]\\ 
\dfrac{1}{\tauin} \left[\psi-\tfrac{p^0\chi\Psi_T}{\sqrt{c_M T^3}}- \dfrac{(\chi, \psi- \tfrac{p^0\chi\Psi_T}{\sqrt{c_M T^3}})}{(\chi, \chi)} \chi \right]+\dfrac{1}{\tauel} (\psi-\langle \psi\rangle)\\
\end{bmatrix} \, .
\end{equation}
One can verify explicitly that both $\E$ and $\s$ are self-adjoint in $\mathcal{H}$.

\newpage

\subsection{Diagonalizing the collision operator}

In order to identify the conserved and non-conserved subspaces, we need to study the spectrum of $\s$. To that end, we notice that any state $\Psi$ can be decomposed as follows:
\begin{equation}\label{decompongo}
\Psi=\underbrace{\dfrac{\Psi_1}{\sqrt{(\chi,\chi)}}
\begin{bmatrix}
0 \\
\chi \\
\end{bmatrix}}_{\Psi_1 e_1} +\underbrace{\dfrac{\Psi_2}{\sqrt{(c_M{+}c_G) T^3}}
\begin{bmatrix}
\sqrt{c_M T^3}\\
\varpi  \\
\end{bmatrix}}_{\Psi_2 e_2}+\underbrace{\dfrac{\Psi_3}{\sqrt{(c_M{+}c_G)T^3}}
\begin{bmatrix}
-\sqrt{c_G T^3} \\
\sqrt{\frac{c_M}{c_G}}\varpi \\ 
\end{bmatrix}}_{\Psi_3 e_3} + 
\begin{bmatrix}
0 \\
\psi_\perp^\text{iso}+\psi^\text{anis}\\
\end{bmatrix} \, ,
\end{equation}
where we have introduced the vector $\varpi=p^0 \chi -\chi \frac{(\chi,p^0\chi)}{(\chi,\chi)}$, which is orthogonal to $\chi$ and has norm $(\varpi,\varpi)=c_G T^3$, with $c_G$ the heat capacity of the gas (see appendix \ref{aaa}). We have also defined $\psi^\text{iso}_\perp=\langle \psi^\text{iso}_\perp \rangle$ to be the isotropic part of $\psi$ orthogonal to $\chi$ and $\varpi$, while $\psi^\text{anis}$ is the purely anisotropic part, such that $\langle\psi^\text{anis}\rangle=0$.

It is immediate to verify that the decomposition \eqref{decompongo} is orthogonal, and that the vectors $e_1$, $e_2$, and $e_3$ are normalized. Moreover, one finds that $\ker(\s)=\text{span}\{e_1,e_2\}$, and
\begin{equation}
\s \Psi
=
\frac{\Psi_3}{\tau_{\mathrm{in}}}
\left(
1{+}\frac{c_G}{c_M}
\right) e_3 +
\begin{bmatrix}
0 \\
\tauin^{-1}\psi_\perp^\text{iso}+(\tauin^{-1}{+}\tauel^{-1})\psi^\text{anis} \\ 
\end{bmatrix} \, .
\end{equation}
Hence, the nonzero spectrum of $\s$ contains three relevant relaxation timescales. The longest one is $\tauin$, which is associated with the relaxation of purely isotropic excitations with vanishing particle and energy densities. These can only relax via inelastic collisions. The anisotropic part relaxes through the action of both elastic and inelastic collisions, and thus over the timescale $\tau=(\tauin^{-1}+\tauel^{-1})^{-1}$. There is also the gas--medium thermal equilibration timescale $\tauT=\tauin/(1+c_G/c_M)$, which corresponds to a situation where the distribution function of the gas is thermal, but its temperature does not agree with that of the medium. In this case, the gas and the medium exchange heat via inelastic collisions, and their temperatures equilibrate. The resulting timescale is shorter than the pure inelastic relaxation time because, while the gas relaxes towards the temperature of the medium, the medium also relaxes towards the temperature of the gas\footnote{Indeed, the value of this third timescale can be derived from a simple thermodynamic argument. Suppose that the gas is thermally distributed with temperature perturbation $\delta T_G$, and that the total energy fluctuation vanishes, $c_M \delta T+c_G\delta T_G=0$. Then, solving the differential equation $\delta \Dot{T}_G=-(\delta T_G-\delta T)/\tauin$, we obtain $\delta T_G\propto e^{-t/\tauT}$, as expected.}.

\subsection{Onsager-normalized density basis}

Given the decomposition \eqref{decompongo}, the natural orthonormal basis of $\mathcal{H}_C$ is $\{e_I\}=\{e_1,e_2\}$. The corresponding conserved densities are $\rho_1=(e_1,\Psi)$ and $\rho_2=(e_2,\Psi)$, which, according to Sec. \ref{quickrecap}, obey
\begin{equation}\label{pdesforrho1andrho2ingeneral}
\partial_t 
\begin{bmatrix}
\rho_1 \\
\rho_2 \\
\end{bmatrix}
=\begin{bmatrix}
\mathfrak{D}_{11} & \mathfrak{D}_{12} \\
\mathfrak{D}_{21} & \mathfrak{D}_{22} \\
\end{bmatrix} \partial_x^2
\begin{bmatrix}
\rho_1 \\
\rho_2 \\
\end{bmatrix}+\text{``Sources''}
\, ,
\end{equation}
where $\mathfrak{D}_{IH}(-\partial_t,\partial_x)$ are the resummed diffusivities \eqref{DIHGeneral} (expressed in real space: $i\omega\to -\partial_t$ and $ik\to \partial_x$).

It is important to keep in mind that the densities $\rho_I$ are Onsager-normalized, meaning that they are linear combinations of $\delta\varepsilon$ and $\delta n$, with coefficients fixed so that their free-energy density is $2\Delta \mathfrak{F}_C=\rho_1^2+\rho_2^2$. In particular, using the definitions of $e_1$ and $e_2$, together with the identities in appendix \ref{aaa}, one finds
\begin{equation}\label{whatrho1andrho2are}
\begin{split}
\rho_1 ={}& \sqrt{\left(\dfrac{\partial \mu}{\partial n} \right)_T} \, \delta n \, , \\
\rho_2={}& \dfrac{\delta\varepsilon-\curlyE \delta n}{\sqrt{(c_M+c_G)T}} \, ,
\end{split}
\end{equation}
where $\curlyE=(\partial\varepsilon_G/\partial n)_T$ is the isothermal energy per added particle of the gas. Here, $\delta\varepsilon=c_M\delta T+\delta\varepsilon_G$ is the perturbation of the total energy density of the medium+gas system.
In the hydrodynamic limit, where the system is in local thermodynamic equilibrium, one has $\delta\varepsilon=(c_M+c_G)\delta T+\curlyE\delta n$, so that $\rho_2$ reduces to $\sqrt{(c_M+c_G)/T}\,\delta T$, and the free-energy perturbation becomes
\begin{equation}
2\Delta \mathfrak{F} \to \left(\dfrac{\partial \mu}{\partial n} \right)_T \, (\delta n)^2+\dfrac{c_M{+}c_G}{T}(\delta T)^2 \, .
\end{equation}
As shown explicitly in appendix \ref{bbb}, for local-equilibrium perturbations of an arbitrary substance, this is precisely the quadratic free-energy perturbation expressed in terms of $\delta n$ and $\delta T$.

\section{Explicit evaluation of the resummed diffusivity matrix}

We have now reached the central part of the paper, where we calculate the resummed diffusivity matrix
\begin{equation}\label{lahardmatrice}
\mathfrak{D}_{IH}(i\omega,ik)= 
\begin{bmatrix}
(\PN\E e_1,[\s_N+ik\E_N-i\omega]^{-1} \PN\E e_1) & (\PN\E e_1,[\s_N+ik\E_N-i\omega]^{-1} \PN\E e_2) \\
(\PN\E e_2,[\s_N+ik\E_N-i\omega]^{-1} \PN\E e_1) & (\PN\E e_2,[\s_N+ik\E_N-i\omega]^{-1} \PN\E e_2)
\end{bmatrix} \, .
\end{equation}
As we shall see, the result admits a fully analytical representation at all frequencies and wave numbers. In particular, no expansion in $\omega$ or $k$ will be required, allowing us to follow the coupled transport of particle number and energy beyond the hydrodynamic regime.

\vspace{-0.3cm}
\subsection{Particle-number diffusivity}
\vspace{-0.3cm}

Since
$\mathcal{H}_C=\text{span}\{e_1,e_2\}$, the relevant projectors read $\PC=e_1(e_1,\cdot)+e_2(e_2,\cdot)$ and $\PN=1-\PC$. Since $e_I$ are isotropic, $\E e_I$ are purely anisotropic and therefore belong to $\mathcal{H}_N$. Hence, we can drop the projectors $\PN$ acting directly on $\E e_I$ in \eqref{lahardmatrice}. Thus, the $11$ entry of \eqref{lahardmatrice} can be written as $\mathfrak{D}_{11}=(\E e_1,\Phi)$, with $\Phi=[\s_N+ik\E_N-i\omega]^{-1} \E e_1$, or, equivalently,
\begin{equation}\label{whatphiis}
[\s_N+ik\E_N-i\omega]\Phi=\E e_1  \, .
\end{equation}
Since $\Phi\in \mathcal{H}_N$ (recall that $\s_N$ and $\E_N$ are operators on $\mathcal{H}_N$), it admits a decomposition
$\Phi=\Phi_3 e_3 +[0,\phi_\perp]^T$, with $\phi_\perp$ orthogonal to $\chi$ and $\varpi$. Hence, \eqref{whatphiis} explicitly reads
\begin{equation}
\Phi_3 [\tauT^{-1}-i\omega+ik \E] e_3+ 
\begin{bmatrix}
0 \\
(\tau^{-1}{-}i\omega{+}ikv^x)\phi_\perp-\tauel^{-1} \langle \phi_\perp \rangle \\
\end{bmatrix}=\dfrac{1}{\sqrt{(\chi,\chi)}}
\begin{bmatrix}
0 \\
v^x\chi \\
\end{bmatrix}+\PC  
\begin{bmatrix}
0 \\
ikv^x\phi_\perp \\
\end{bmatrix}  \, .
\end{equation}
By examining the first row of this equation, we obtain
\begin{equation}\label{thePhi3}
\Phi_3 =-\dfrac{ik(\varpi,v^x \phi_\perp)}{(\tauT^{-1}-i\omega)\sqrt{(1{+}\frac{c_G}{c_M})c_G T^3}}  \, .
\end{equation}
The second component then gives
\begin{equation}\label{themonsterequation}
\phi_\perp =\dfrac{\tauel^{-1} \langle \phi_\perp \rangle+\dfrac{v^x\chi}{\sqrt{(\chi,\chi)}}+ik\dfrac{\chi(
\chi,v^x\phi_\perp)}{(\chi,\chi)}+ik\left[\dfrac{c_G}{c_M} +\dfrac{\tauT^{-1}{-}i\omega{+}ik v^x}{\tauT^{-1}{-}i\omega  }\right]\dfrac{\varpi(\varpi ,  
v^x\phi_\perp)}{(1{+}\frac{c_G}{c_M})c_G T^3} }{\tau^{-1}{-}i\omega{+}ikv^x} \, .
\end{equation}
Hence, defining the functions
\begin{equation}
\begin{split}
H_0(i\omega,ik)\equiv{}& \left\langle \dfrac{1}{\tau^{-1}{-}i\omega{+}ikv^x}\right\rangle= \dfrac{1}{ik} \text{arctanh}\left(\dfrac{ik\tau}{1{-}i\omega\tau} \right) \, ,\\
H_1(i\omega,ik)\equiv{}& \left\langle \dfrac{v^x}{\tau^{-1}{-}i\omega{+}ikv^x}\right\rangle= \dfrac{1}{ik} \left[1-\dfrac{1{-}i\omega\tau}{ik\tau} \text{arctanh}\left(\dfrac{ik\tau}{1{-}i\omega\tau} \right)\right] \, ,\\
H_2(i\omega,ik)\equiv{}& \left\langle \dfrac{(v^x)^2}{\tau^{-1}{-}i\omega{+}ikv^x}\right\rangle= -\dfrac{1{-}i\omega\tau}{(ik)^2\tau} \left[1-\dfrac{1{-}i\omega\tau}{ik\tau} \text{arctanh}\left(\dfrac{ik\tau}{1{-}i\omega\tau} \right)\right] \, ,\\
\end{split}
\end{equation}
we can average \eqref{themonsterequation} over all angles, and we obtain
\begin{equation}\label{theisotorpicmonsterequation}
\langle\phi_\perp\rangle =\dfrac{H_1}{1{-}H_0 \tauel^{-1}}\dfrac{\chi}{\sqrt{(\chi,\chi)}}+\dfrac{ikH_0}{1{-}H_0 \tauel^{-1}}\dfrac{\chi(
\chi,v^x\phi_\perp)}{(\chi,\chi)}+\dfrac{ikH_0}{1{-}H_0 \tauel^{-1}}\left[\dfrac{c_G}{c_M} +\dfrac{\tauT^{-1}{-}i\omega{+}ik H_1/H_0}{\tauT^{-1}{-}i\omega  }\right]\dfrac{\varpi(\varpi ,  
v^x\phi_\perp)}{(1{+}\frac{c_G}{c_M})c_G T^3}  \, .
\end{equation}
If now we take the inner product of \eqref{theisotorpicmonsterequation} with $\chi$ and $\varpi$, and use $0=(\chi,\phi_\perp)=(\chi,\langle\phi_\perp\rangle)$ and $0=(\varpi,\phi_\perp)=(\varpi,\langle\phi_\perp\rangle)$, where the second equality in each relation follows from the isotropy of $\chi$ and $\varpi$, we find
\begin{equation}
\begin{split}
(\chi,v^x\phi_\perp) ={}& -\dfrac{H_1}{ikH_0} \sqrt{(\chi,\chi)} \, , \\
(\varpi ,  
v^x\phi_\perp) ={}& 0 \, . \\
\end{split}
\end{equation}
Plugging these expressions back into \eqref{thePhi3}, \eqref{theisotorpicmonsterequation}, and \eqref{themonsterequation}, we find that $\Phi_3=0$, $\langle \phi_\perp \rangle=0$, and
\begin{equation}
\phi_\perp = \dfrac{v^x-H_1/H_0 }{\tau^{-1}{-}i\omega{+}ikv^x} \dfrac{\chi}{\sqrt{(\chi,\chi)}} \, .    
\end{equation}
One can verify explicitly that this $\Phi$ does indeed solve equation \eqref{whatphiis}.
\newpage

Now that we know $\Phi$, we only need to compute the inner product $(\E e_1,\Phi)$ to determine the $11$ diffusivity. It is straightforward to verify that $\mathfrak{D}_{11}=H_2-H_1^2/H_0$, which simplifies to
\begin{equation}\label{D11explicit}
\mathfrak{D}_{11}(i\omega,ik)=\dfrac{(1-i\omega\tau)\tau}{(ik\tau)^2}\left[1-\dfrac{\frac{ik\tau}{1-i\omega\tau}}{\text{arctanh}\left(\frac{ik\tau}{1-i\omega\tau}\right)}\right] \, .
\end{equation}
The knowledge of $\Phi$ also allows us to compute the off-diagonal diffusivity $\mathfrak{D}_{21}=(\E e_2,\Phi)$, while the identity $\mathfrak{D}_{12}(i\omega,ik)=\mathfrak{D}^*_{21}((i\omega)^*,(ik)^*)$ determines $\mathfrak{D}_{12}$. Remarkably, we find
\begin{equation}\label{D12D21arezero}
\mathfrak{D}_{12}(i\omega,ik)=\mathfrak{D}^*_{21}((i\omega)^*,(ik)^*)=0 \, .
\end{equation}
Thus, our choice of basis $e_I$ is particularly natural for this problem, since it diagonalizes not only the free energy, but also the resummed diffusivity matrix. At $i\omega=ik=0$, such a simultaneous diagonalization is always possible in an Onsager-normalized basis. However, there is no guarantee in general that $\mathfrak{D}_{IH}$ will remain diagonal in the same basis at finite $i\omega$ and $ik$. In our case, this happens because the dynamics preserves the separation between the two radial structures $\chi$ and $\varpi$: indeed, $\phi_\perp=\chi\times\text{``purely angular function''}$, while $\E e_2=\varpi\times\text{``purely angular function''}$. Since $(\chi,\varpi)=0$, the two sectors remain orthogonal at all frequencies and wave numbers. This separation would generally be lost if the particle velocity or the relaxation times depended on the particle energy.

\vspace{-0.3cm}
\subsection{Thermal diffusivity}
\vspace{-0.3cm}

The $22$ entry of the diffusivity matrix is calculated following a procedure analogous to the one above. This time, we have that $\mathfrak{D}_{22}=(\E e_2,\Xi)$, where the vector $\Xi\in \mathcal{H}_N$ solves the equation
\begin{equation}\label{theluixi}
[\s_N+ik\E_N-i\omega]\Xi =\E e_2 \, .
\end{equation}
As before, we have a decomposition $\Xi=\Xi_3 e_3+[0,\xi_\perp]^T$, with $\xi_\perp$ orthogonal to both $\chi$ and $\varpi$. This time, however, there is a useful simplification. Since we have already shown that $\mathfrak{D}_{12}=0$, we know that $0=\mathfrak{D}_{12}=(\E e_1,\Xi)$, or, equivalently, $(\chi, v^x \xi_\perp)=0$.
Therefore, equation \eqref{theluixi} explicitly reads
\begin{equation}
\begin{bmatrix}
0 \\
(\tau^{-1}{-}i\omega{+}ikv^x)\xi_\perp{-}\tauel^{-1}\langle \xi_\perp\rangle \\
\end{bmatrix}=\dfrac{1}{\sqrt{(c_M{+}c_G) T^3}}
\begin{bmatrix}
0\\
v^x\varpi  \\
\end{bmatrix}+\dfrac{ik(\varpi,v^x\xi_\perp)}{(c_M{+}c_G) T^3}
\begin{bmatrix}
\sqrt{c_M T^3}\\
\varpi  \\
\end{bmatrix}+\dfrac{\Xi_3 (\tauT^{-1}{-}i\omega{+}ik\E)}{\sqrt{(c_M{+}c_G)T^3}}
\begin{bmatrix}
\sqrt{c_G T^3} \\
-\sqrt{\frac{c_M}{c_G}}\varpi \\ 
\end{bmatrix} \! .
\end{equation}
As before, if we analyze the two rows one at a time, and average over all angles, we obtain
\begin{equation}\label{epfmerfp}
\begin{split}
\Xi_3={}&-\dfrac{ik(\varpi,v^x\xi_\perp)}{(\tauT^{-1}-i\omega)\sqrt{(1{+}\frac{c_G}{c_M}) c_G T^3}} \, ,\\
\xi_\perp ={}&\dfrac{\tauel^{-1}\langle \xi_\perp\rangle+\dfrac{v^x\varpi}{\sqrt{(c_M{+}c_G) T^3}} +ik\left[\dfrac{c_G}{c_M} +\dfrac{\tauT^{-1}{-}i\omega{+}ik v^x}{\tauT^{-1}{-}i\omega}\right]\dfrac{\varpi (\varpi,v^x\xi_\perp)}{(1{+}\frac{c_G}{c_M})c_G T^3}}{\tau^{-1}{-}i\omega{+}ikv^x},\\
\langle\xi_\perp\rangle ={}&\dfrac{H_1}{1- H_0\tauel^{-1}}\dfrac{\varpi}{\sqrt{(c_M{+}c_G) T^3}} +\dfrac{ikH_0}{1- H_0\tauel^{-1}}\left[\dfrac{c_G}{c_M} +\dfrac{\tauT^{-1}{-}i\omega{+}ik H_1/H_0}{\tauT^{-1}{-}i\omega}\right]\dfrac{\varpi (\varpi,v^x\xi_\perp)}{(1{+}\frac{c_G}{c_M})c_G T^3}.\\
\end{split}
\end{equation}
The third equation implies that $\langle\xi_\perp\rangle\propto \varpi$. But since $\langle\xi_\perp\rangle$ is orthogonal to $\varpi$ (recall that $0=(\varpi,\xi_\perp)=(\varpi,\langle\xi_\perp\rangle)$, where the second equality follows from the isotropy of $\varpi$), we can only have $\langle\xi_\perp\rangle=0$. This gives 
\begin{equation}
(\varpi,v^x\xi_\perp) =-\dfrac{H_1}{ikH_0}\dfrac{\sqrt{(c_M{+}c_G) T^3}}{1 +\dfrac{c_M(\tauT^{-1}{-}i\omega{+}ik H_1/H_0)}{c_G(\tauT^{-1}{-}i\omega)}} \, ,
\end{equation}
and since $\mathfrak{D}_{22}=(\E e_2,\Xi)=\frac{1}{\sqrt{(c_M+c_G)T^3}}(\varpi,v^x\xi_\perp)$, we finally obtain
\begin{equation}\label{D22explicit}
 \mathfrak{D}_{22}(i\omega,ik)=  \dfrac{\mathfrak{D}_{11}(i\omega,ik)}{1 +\dfrac{c_M}{c_G}\left[1-\dfrac{\mathfrak{D}_{11}(i\omega,ik)(ik)^2\tauT}{1{-}i\omega\tauT}\right]} 
\end{equation}
Unlike $\mathfrak{D}_{11}$, which depends on the collision times only through the total relaxation time $\tau$, the thermal diffusivity depends explicitly on $\tauT$ and $c_M/c_G$.

\subsection{Structure of the resummed diffusivities}
\vspace{-0.3cm}

We now discuss the physical content of the diffusivities obtained above. 

Let us start from the particle-number sector. In the homogeneous limit, we find
\begin{equation}\label{halken}
\mathfrak{D}_{11}(i\omega,0)=\dfrac{\tau}{3(1-i\omega\tau)} \, .
\end{equation}
At zero frequency, this reduces to the familiar random-walk result $\mathfrak{D}_{11}(0,0)=\tau/3$, for particles moving at unit speed in three spatial dimensions. At finite frequency, the additional factor $(1-i\omega\tau)^{-1}$ is the usual Cattaneo correction, encoding the finite relaxation time of the particle current. At finite wave number, however, \eqref{D11explicit} contains genuinely kinetic information: the inverse hyperbolic tangent generates a branch cut connecting $\omega=-k-i/\tau$ and $\omega=k-i/\tau$, corresponding to the continuum of non-hydrodynamic modes $\omega=kv^x-i/\tau$. Thus, the resummed diffusivity captures the continuous kinetic spectrum that is invisible to any finite-order hydrodynamic expansion.

The thermal diffusivity has a richer structure. At $k=0$, we find
\begin{equation}\label{chaskiort}
\mathfrak{D}_{22}(i\omega,0)=\dfrac{c_G}{c_M+c_G}\dfrac{\tau}{3(1-i\omega\tau)} \, .
\end{equation}
Hence, the same random-walk result is recovered, but suppressed by the fraction of the total heat capacity carried by the mobile gas \cite{GavassinoRadiativeBounds:2025bxx}. Indeed, both the gas and the medium store thermal energy, while only the gas can transport it. The dependence on $\tauT$ drops out once again because of the angular structure of the modes. In fact, from \eqref{lahardmatrice}, we have that $\mathfrak{D}_{IH}(i\omega,0)=(\E e_I,[\s_N-i\omega]^{-1}\E e_H)$. Since the vectors $\E e_H$ are purely anisotropic, they relax on the timescale $\tau$, and therefore $\mathfrak{D}_{IH}(i\omega,0)=(\tau^{-1}-i\omega)^{-1}(\E e_I,\E e_H)$, in agreement with \eqref{halken} and \eqref{chaskiort}.
At finite wave number, this argument no longer applies, and $\tauT$ enters \eqref{D22explicit} explicitly. 

\begin{figure}[h!]
    \centering
\includegraphics[width=0.46\linewidth]{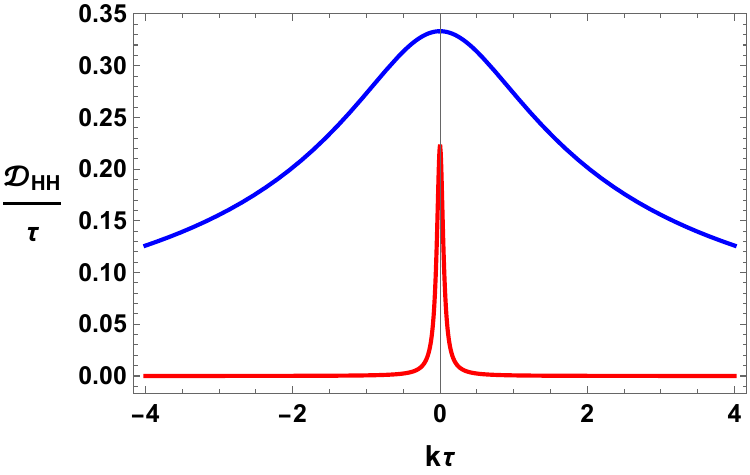}\hspace{0.5cm}
\includegraphics[width=0.46\linewidth]{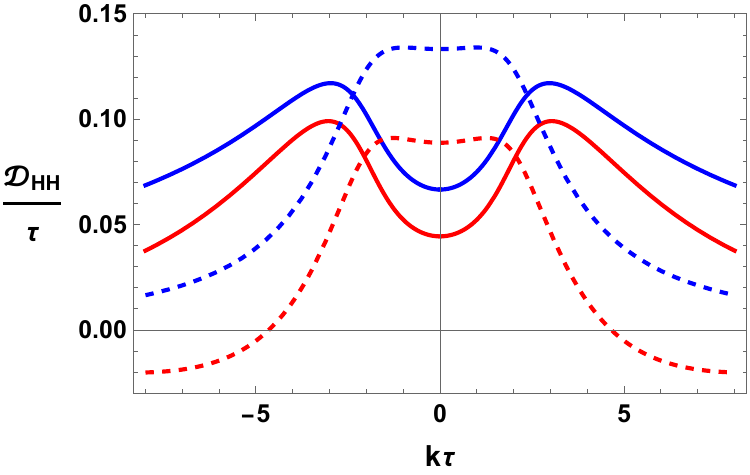}
\caption{Resummed particle-number (blue) and thermal (red) diffusivities as functions of real $k$, for $c_M/c_G{=}0.5$ and $\tauin/\tauel{=}10^4$. \textit{Left panel}: static diffusivities ($\omega=0$), both of which are real. At $k=0$, the thermal current only involves anisotropic distortions of the gas distribution, which relax on the timescale $\tau$, so that $\tauT$ plays no role. At finite $k$, instead, spatial transport locally drives the gas and the stationary medium out of thermal equilibrium. Since this temperature imbalance relaxes on the much longer timescale $\tauT$, it persists and strongly suppresses the transport of total thermal energy, causing $\mathfrak{D}_{22}$ to decay rapidly away from $k=0$. \textit{Right panel}: real (solid) and imaginary (dashed) parts at $\omega\tau=2$. When $\omega\tauT\gg1$, the gas--medium temperature imbalance cannot relax within one oscillation, and its response is controlled by the driving frequency rather than by $\tauT$. The singular sharpening of $\mathfrak{D}_{22}$ with increasing $\tauT$ is therefore cut off.}
    \label{fig:DHHexplicit}
\end{figure}

Taking $\tau$ as the reference microscopic timescale, we have
\begin{equation}\label{tauTovertau}
\dfrac{\tauT}{\tau}
=
\dfrac{1+\tauin/\tauel}{1+c_G/c_M} \, .
\end{equation}
Thus, $\tauT/\tau$ can be small only when $c_M/c_G$ is small. In this limit, the heat capacity of the stationary medium becomes negligible, and \eqref{D22explicit} gives $\mathfrak{D}_{22}\rightarrow\mathfrak{D}_{11}$. The opposite limit is more interesting. At fixed heat capacities, $\tauT/\tau$ can become arbitrarily large by taking $\tauin/\tauel\rightarrow\infty$. In the static limit, the factor $\tauT/(1-i\omega\tauT)$ in \eqref{D22explicit} reduces to $\tauT$, and grows without bound. As a result, while $\mathfrak{D}_{22}(0,0)=c_G\tau/[3(c_M+c_G)]$ remains finite, $\mathfrak{D}_{22}(0,ik)$ approaches zero for any fixed real $k\neq0$, producing an increasingly narrow peak around $k=0$ (as shown in Fig.~\ref{fig:DHHexplicit}). This singular behavior is cut off at finite real frequency, since $\tauT/(1-i\omega\tauT)\rightarrow i/\omega$ as $\tauT\rightarrow\infty$. Thus, at nonzero frequency, increasing $\tauT$ no longer produces an arbitrarily sharp dependence on $k$.

\subsection{Coupled transport and the Onsager matrix}

Although $\rho_1$ and $\rho_2$ diffuse independently, particle and energy transport are still coupled. In the conventional description of thermoelectric transport, this coupling is encoded in the Onsager matrix, whose off-diagonal entries describe the particle current induced by a temperature gradient and the energy current induced by a chemical-potential gradient. We now verify explicitly that the diagonal form of the resummed diffusivity matrix is fully consistent with the existence of this coupling. 

First of all, let us note that, since $\rho_1\propto \delta n$, and $\rho_2\propto \delta\varepsilon-\curlyE \delta n$ (see equation \eqref{whatrho1andrho2are}), the diffusivity matrix in the basis $\{\delta n,\delta\varepsilon\}$ has an off-diagonal term:
\begin{equation}\label{physicaldiffusionequations}
\partial_t
\begin{bmatrix}
\delta n\\
\delta\varepsilon
\end{bmatrix}
=
\begin{bmatrix}
\mathfrak D_{11} & 0\\
\curlyE(\mathfrak D_{11}-\mathfrak D_{22}) & \mathfrak D_{22}
\end{bmatrix}
\partial_x^2
\begin{bmatrix}
\delta n\\
\delta\varepsilon
\end{bmatrix}.
\end{equation}
Hence, a particle-density gradient generically contributes to the energy current. The off-diagonal diffusivity only vanishes in the limit $\mathfrak D_{22}\to\mathfrak D_{11}$, which, we recall, is attained when $c_M\ll c_G$ (i.e. in gas-dominated systems). As we shall see, however, the thermoelectric coupling in the Onsager matrix survives even in this limit.

To recover the conventional Onsager matrix \cite{Onsager_1931,Onsager_Casimir,landau5}, we take the hydrodynamic limit $\omega,k\to0$, so that local thermodynamic equilibrium can be assumed. Then, the thermodynamic relations of Appendix \ref{aaa} (taking the medium into account) give
\begin{equation}
\begin{split}
\delta n &=
\beta (\chi,\chi)
\left[\delta(\mu/T)-\curlyE\,\delta(1/T)\right],\\
\delta\varepsilon &=
\curlyE \beta (\chi,\chi)\delta(\mu/T)
-\left[
\curlyE^2\beta(\chi,\chi)
+(c_M+c_G)T^2
\right]\delta (1/T) .
\end{split}
\end{equation}
Writing the conservation laws as $\partial_t\delta n+\partial_x\delta J_n=0$ and $\partial_t\delta\varepsilon+\partial_x \delta J_\varepsilon=0$, equation \eqref{physicaldiffusionequations} then takes the standard Onsager form
\begin{equation}\label{Onsagermatrixphysical}
\begin{bmatrix}
\delta J_n\\
\delta J_\varepsilon
\end{bmatrix}
=-
\begin{bmatrix}
L_{11} & L_{12}\\
L_{21} & L_{22}
\end{bmatrix}\partial_x
\begin{bmatrix}
\delta(\mu/T)\\
-\delta(1/T)
\end{bmatrix},
\end{equation}
where
\begin{equation}\label{Onsagermatrixexplicit}
\begin{bmatrix}
L_{11} & L_{12}\\
L_{21} & L_{22}
\end{bmatrix}
=
\begin{bmatrix}
\beta(\chi,\chi) \mathfrak{D}_{11}
& &
\curlyE \beta(\chi,\chi) \mathfrak{D}_{11}
\\[2mm]
\curlyE \beta(\chi,\chi) \mathfrak{D}_{11}
& &
\curlyE^2\beta(\chi,\chi) \mathfrak{D}_{11}
+(c_M{+}c_G)T^2\mathfrak{D}_{22}
\end{bmatrix}_{\omega=k=0}.
\end{equation}
Note that $L_{12}=L_{21}$ due to Onsager reciprocity.

Using \eqref{halken}, \eqref{chaskiort}, and \eqref{latermo}, the above matrix reduces to
\begin{equation}
\begin{bmatrix}
L_{11} & L_{12}\\
L_{21} & L_{22}
\end{bmatrix}
=
\frac{T\tau}{3}
\begin{bmatrix}
\left(\dfrac{\partial n}{\partial\mu}\right)_T
& &
\curlyE\left(\dfrac{\partial n}{\partial\mu}\right)_T
\\[2mm]
\curlyE\left(\dfrac{\partial n}{\partial\mu}\right)_T
& &
\curlyE^2\left(\dfrac{\partial n}{\partial\mu}\right)_T{+}c_G T
\end{bmatrix}.
\end{equation}
The complete disappearance of $c_M$ from the Onsager matrix is not accidental. Although the medium contributes to the total energy density, it does not carry a current; accordingly, only the heat capacity $c_G$ of the mobile gas survives in the first-order transport coefficients.

The off-diagonal coefficients in the Onsager matrix have the usual thermoelectric interpretation \cite[\S 26]{landau8}. The coefficient $L_{12}$ describes the particle current generated by a temperature gradient, while $L_{21}$ describes the energy current generated by a chemical-potential gradient. For charged mobile particles, these give the usual Seebeck and Peltier effects, respectively. In the present model, both effects originate from the fact that the mobile particles carry an average isothermal energy $\curlyE$ per added particle, which explains the common factor $\curlyE$ in the two cross coefficients. The corresponding Seebeck coefficient can be obtained explicitly by imposing $\delta J_n\,{=}\,0$. From \eqref{Onsagermatrixphysical}, this gives $\partial_x(\mu/T)\,{=}\,\curlyE\,\partial_x(1/T)$, and therefore
\begin{equation}
\left.\frac{\delta\mu}{\delta T}\right|_{\delta J_n=0}
=-\frac{\curlyE-\mu}{T}\, .
\end{equation}
Thus, setting the particle charge to unity, the Seebeck coefficient is precisely $S=(\curlyE-\mu)/T$. This agrees with the standard result obtained from the Boltzmann equation in the relaxation-time approximation, $S=\mathcal L^{(1)}/[T\mathcal L^{(0)}]$ \cite{MadsenSingh2006,MadsenBoltzTraP22018}. Indeed, for ultrarelativistic particles and energy-independent relaxation times, the velocity and relaxation-time factors cancel from the ratio of transport moments, yielding $\mathcal L^{(1)}/\mathcal L^{(0)}=\curlyE-\mu$. Thus, even though the Onsager-normalized densities $\rho_1$ and $\rho_2$ diffuse independently, particle and energy currents exhibit the usual thermoelectric cross responses.

\section{Dispersion relations}

We now use the resummed diffusivities to determine the dispersion relations of the hydrodynamic modes, together with the isolated non-hydrodynamic modes that carry nonzero conserved densities. Since $\rho_1$ and $\rho_2$ decouple, their dispersion relations can be determined independently from the corresponding conservation laws without source terms.

\subsection{Particle diffusion}

Combining equations \eqref{pdesforrho1andrho2ingeneral}, \eqref{whatrho1andrho2are}, and \eqref{D12D21arezero}, we immediately see that the conservation law for $\rho_1$ is equivalent to
\begin{equation}
\partial_t \delta n =\mathfrak{D}_{11}(-\partial_t ,\partial_x )\partial^2_x\delta n \qquad \qquad (\partial_t \to -i\omega, \, \, \partial_x \to ik) \, .
\end{equation}
Due to the simplicity of \eqref{D11explicit}, this equation can be solved analytically, giving
\begin{equation}\label{RTAone}
i\omega\tau =1- \dfrac{k\tau}{\tan(k\tau)}\, , \qquad \qquad k\tau\in \left[-\dfrac{\pi}{2}, \dfrac{\pi}{2} \right] \, ,
\end{equation}
which is identical to the dispersion relation of particle diffusion in ultrarelativistic kinetic theory in the Relaxation-Time Approximation (RTA) \cite{RomatschkeCutsandPoles:2015gic}. Physically, this tells us that, if we prepare a local density excess in the gas at the background temperature, the subsequent evolution preserves this temperature: neither $e_2$ nor $e_3$ is excited, and all non-equilibrium corrections are purely anisotropic. Elastic and inelastic scatterings then act identically, entering the dynamics only through the total relaxation time $\tau$.

This can also be seen directly from the equations of motion. In particular, if we make the ansatz $\delta T=0$ and $\delta f=\feq(1+s\feq)I(v^j)$, with $I\in L^2(S^2)$, equation \eqref{matchingconditiondeltaalpha} gives $\delta\alpha=\langle I\rangle$, while system \eqref{kineticmodel} reduces to
\begin{equation}\label{kineticmodelAnsatzed}
(\partial_t {+}v^x\partial_x) I  =\dfrac{\langle I\rangle -I}{\tau}  \, .
\end{equation}
This is precisely the RTA kinetic equation leading to \eqref{RTAone}, proving the consistency of the ansatz. Note that the excitation $\delta f=\feq(1+s\feq)I$ carries both a finite particle density and a finite energy density, but these are proportional to each other via $\delta\varepsilon=\curlyE\delta n$. From \eqref{whatrho1andrho2are}, this is precisely the condition for a vanishing $\rho_2$.

\subsection{Thermal diffusion}

\begin{figure}[h!]
    \centering
\includegraphics[width=0.45\linewidth]{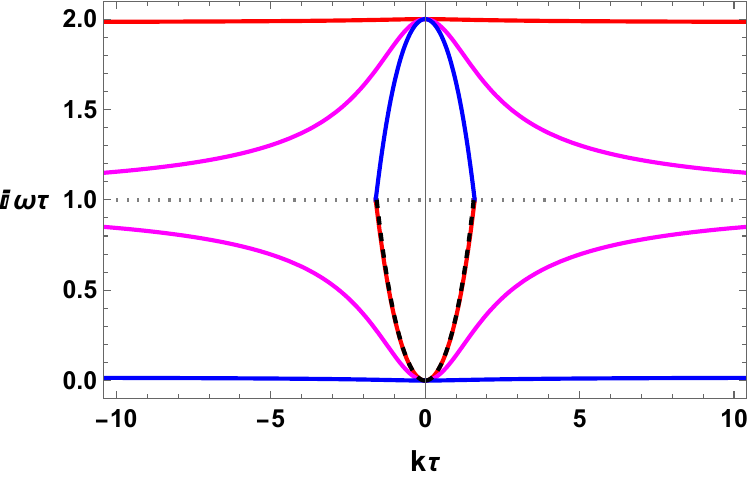}\hspace{0.5cm}
\includegraphics[width=0.45\linewidth]{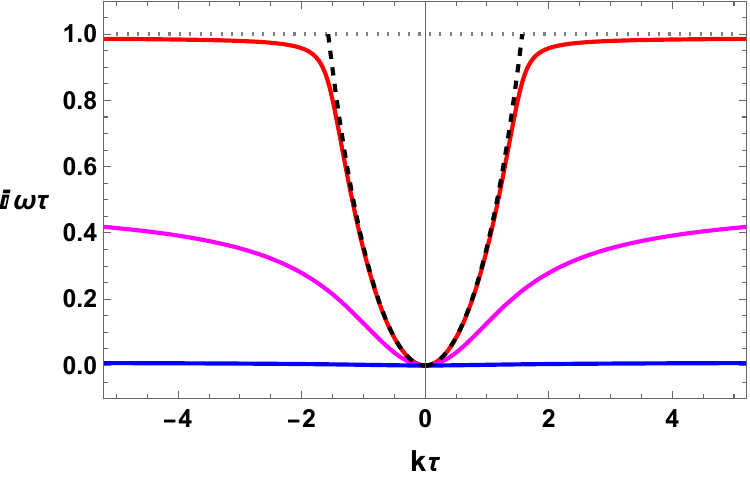}
\includegraphics[width=0.45\linewidth]{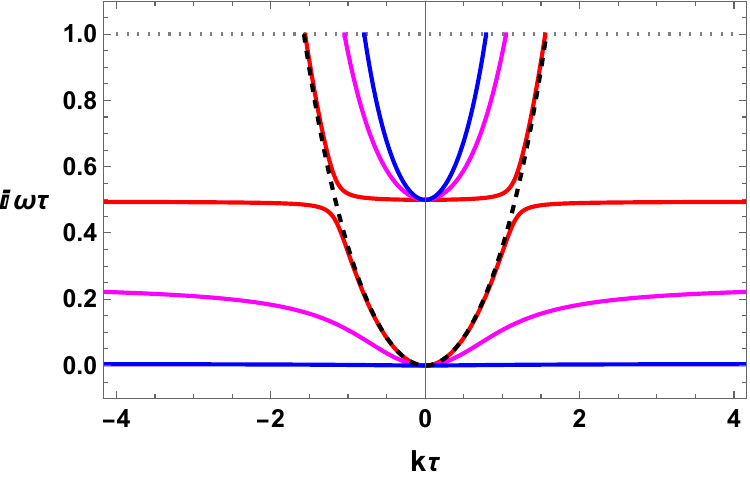}\hspace{0.5cm}
\includegraphics[width=0.45\linewidth]{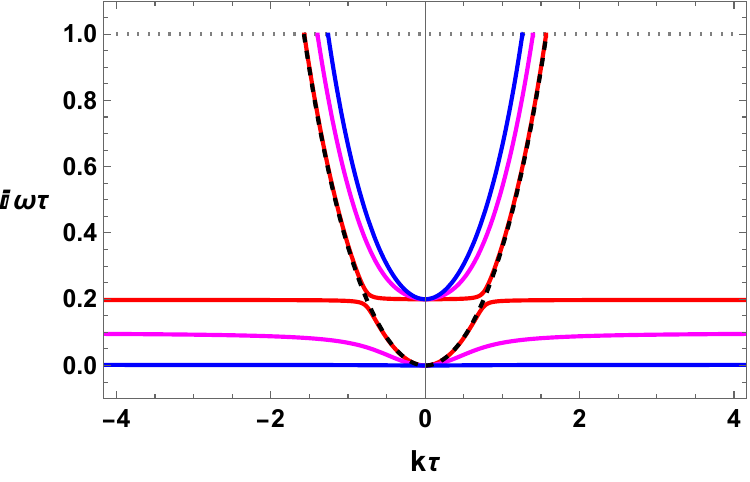}
\caption{Dispersion relations in the thermal sector for $\tauT/\tau=1/2$ (upper left), $1$ (upper right), $2$ (lower left), and $5$ (lower right). The red, magenta, and blue curves correspond respectively to $c_M/c_G=0.01$, $1$, and $100$. Here $\tauT/\tau$ and $c_M/c_G$ are treated as independent parameters to display more clearly the different regimes, even though not all combinations are compatible with \eqref{tauTovertau}. The black dashed curve is the particle-diffusion dispersion relation \eqref{RTAone}, while the horizontal gray dotted line marks the branch cut at $i\omega\tau=1$, where isolated modes merge into the ballistic continuum. Each color contains a hydrodynamic mode starting from $\omega=0$ and a non-hydrodynamic mode starting from $i\omega=1/\tauT$. When the heat capacity is dominated by the mobile gas, $c_M/c_G\ll1$ (red), thermal diffusion initially follows almost exactly the particle-diffusion curve, since essentially all thermal energy is stored in the particles that transport it. If $\tauT>\tau$, however, the hydrodynamic decay rate eventually saturates at $i\omega\simeq1/\tauT$: energy stored in the stationary medium must be transferred to the gas before it can be transported, so thermal equilibration becomes the bottleneck. At the same time, the non-hydrodynamic mode takes on the character of diffusive transport, so that the two modes effectively ``exchange'' roles. In the opposite limit $c_M/c_G\gg1$ (blue), most of the energy is stored in the stationary medium while only a negligible fraction is carried by the gas; consequently, the thermal diffusivity tends to zero and the hydrodynamic mode remains arbitrarily close to $i\omega=0$. The case $c_M/c_G=1$ (magenta) interpolates between these two regimes.}
    \label{fig:Disp}
\end{figure}

Out of local equilibrium, temperature is not uniquely defined, due to hydrodynamic frame ambiguities \cite{Kovtun2019}. We define a non-equilibrium temperature by $\delta T_{DF}=(\delta\varepsilon{-}\curlyE\delta n)/(c_M{+}c_G)\propto\rho_2$, in analogy with the density frame \cite{ArmasDensityFrame2020mpr,BhambureDensityFrame2024axa}. Hence, combining \eqref{pdesforrho1andrho2ingeneral} with \eqref{D12D21arezero}, we find
\begin{equation}
\partial_t \delta T_{DF} =\mathfrak{D}_{22}(-\partial_t ,\partial_x )\partial^2_x\delta T_{DF} \qquad \qquad (\partial_t \to -i\omega, \, \, \partial_x \to ik) \, .
\end{equation}
Also in this case, the dispersion relations can be obtained explicitly, although only in parametric form. In particular, defining $r=k\tau/(1-i\omega\tau)$, we have $i\omega\tau=\Gamma(r)$ and $k\tau=r[1-\Gamma(r)]$, where $\Gamma(r)$ satisfies
\begin{equation}\label{pieroski}
\dfrac{(\Gamma-1) r}{\arctan(r)} \left[\left(1+\frac{c_M}{c_G}\right)\frac{\tauT}{\tau}\Gamma-1\right]+ \frac{c_M}{c_G} \left(\frac{\tauT}{\tau}-1\right)\Gamma+ \frac{\tauT}{\tau}\Gamma-1=0\, .
\end{equation}
There are in general two roots, corresponding to the hydrodynamic mode and to the non-hydrodynamic mode emanating from $i\omega=1/\tauT$. These are shown in figure \ref{fig:Disp} for several choices of the parameters.

The thermal sector can also be decoupled directly at the level of the equations of motion. In particular, setting $\delta f=\feq(1+s\feq)(p^0-\curlyE)\beta^2 I(v^j)$, we have $\rho_1\propto\delta n=0$, while $\delta\alpha=-\curlyE\beta^2\delta T$, and \eqref{kineticmodel} reduces to
\begin{equation}\label{kineticmodelansatzed2}
\begin{split}
& c_M \partial_t \delta T +\dfrac{c_G}{\tauin} [\delta T -\langle I\rangle]   =0 \, , \\
& (\partial_t {+}v^x\partial_x) I  =\dfrac{\delta T -I}{\tauin} +\dfrac{\langle I\rangle -I}{\tauel}  \, ,
\end{split}
\end{equation}
whose dispersion relations are precisely the solutions of \eqref{pieroski}.

\newpage
\section{Conclusions}

In this work, we have presented an analytically solvable example of a system with two conserved diffusing densities for which the full matrix of resummed diffusivities can be obtained in closed form. The model describes ultrarelativistic particles transporting particle number and energy through a rigid material medium, with elastic collisions randomizing their direction and inelastic collisions exchanging energy with a dynamical thermal background. Despite its tractability, the model combines two distinct scattering mechanisms, gas--medium thermal equilibration, and general quantum statistics, and therefore captures a considerably richer transport structure than standard relaxation-time models.

The Onsager-normalized conserved densities $\rho_1$ and $\rho_2$ diagonalize the diffusivity matrix not only in the hydrodynamic limit, but at arbitrary frequency and wave number. The resulting diffusivities resum the constitutive derivative expansion to all orders. While $\mathfrak D_{11}$ depends on the microscopic collisions only through the total relaxation time $\tau$, which receives contributions from both elastic and inelastic scattering, the thermal diffusivity $\mathfrak D_{22}$ additionally resolves the gas--medium equilibration time $\tauT$. This scale is invisible in the Fick limit but emerges at finite wave number, where streaming can drive the mobile gas out of local thermal equilibrium with the stationary medium. The resummed diffusivities thus expose microscopic information lost in first-order hydrodynamics, while also retaining genuinely kinetic features such as the ballistic continuum encoded in their branch cuts.

The diagonal form of the diffusivity matrix in the $(\rho_1,\rho_2)$ basis does not imply decoupled physical transport. In terms of particle number and total energy, the diffusion equations are generically coupled, and their hydrodynamic limit reproduces a fully coupled Onsager matrix with the expected thermoelectric cross coefficients. In particular, the Seebeck coefficient agrees with the standard relaxation-time kinetic-theory result \cite{MadsenSingh2006,MadsenBoltzTraP22018}. Thus, the same exact solution simultaneously captures diffusion, thermoelectric coupling, thermal equilibration, and non-hydrodynamic kinetic dynamics.

More generally, the resummed diffusivities contain strictly more information than the dispersion relations of the hydrodynamic modes. Dispersion relations follow only after imposing the source-free conservation laws and therefore probe the system on shell. By contrast, $\mathfrak D_{IH}(i\omega,ik)$ defines the constitutive relation between densities and currents also in the presence of external sources, determining the linear response at arbitrary $\omega$ and $k$ \cite{GavassinoDiamond:2026fgz}. The dispersion relations can therefore be reconstructed from the resummed diffusivities, whereas the converse is not possible.

Closed-form resummed diffusivities will generally be unavailable in more realistic microscopic theories, but the operator formulation provides a systematic route forward. As shown in Ref.~\cite{GavassinoDiamond:2026fgz}, the relevant resolvent admits a convergent Neumann expansion around the hydrodynamic limit, allowing off-shell constitutive relations to be constructed directly from microscopic collision and streaming operators. The present model provides an exact benchmark for this broader program and illustrates the information gained by resumming constitutive relations rather than individual dispersion relations. This perspective is particularly promising for multicomponent kinetic systems with several coupled conserved charges, including neutrino transport in dense matter and the simultaneous diffusion of baryon number, strangeness, and electric charge (BSQ) in relativistic nuclear matter, where multiple microscopic relaxation channels and coupled diffusive modes arise naturally.

\vspace{-0.3cm}
\section*{Acknowledgements}
\vspace{-0.3cm}

This work is supported by a MERAC Foundation prize grant,  an Isaac Newton Trust Grant, and funding from the Cambridge Centre for Theoretical Cosmology.

\appendix
\vspace{-0.3cm}
\section{Thermodynamic susceptibilities as inner products}\label{aaa}
\vspace{-0.3cm}

In this appendix, we show how inner products involving $\chi$ and $\varpi$ can be interpreted as thermodynamic susceptibilities of the gas. In the following, it is understood that all thermodynamic quantities refer to the gas alone.

In thermodynamic equilibrium, the distribution function of the gas is $\feq(\alpha,T)=(e^{-\alpha+p^0/T}-s)^{-1}$, while the particle and energy densities are
\begin{equation}
n=\int \dfrac{gd^3p}{(2\pi)^3} \feq \, , \qquad\qquad \varepsilon=\int \dfrac{gd^3p}{(2\pi)^3} p^0\feq \, .
\end{equation}
Differentiating with respect to $T$ and $\alpha$, and identifying $\feq$ with the background distribution entering \eqref{definechi}, we obtain the identities
\begin{equation}\label{latermo}
\begin{split}
\left(\dfrac{\partial n}{\partial \alpha}\right)_T={}& \beta(\chi,\chi) \, , \\
\left(\dfrac{\partial n}{\partial T}\right)_\alpha={}& \beta^3(\chi,p^0\chi) \, , \\
\left(\dfrac{\partial \varepsilon}{\partial \alpha}\right)_T={}& \beta(\chi,p^0\chi) \, , \\
\left(\dfrac{\partial \varepsilon}{\partial T}\right)_\alpha={}& \beta^3(p^0\chi,p^0\chi) \, . \\
\end{split}
\end{equation}
Thus, using Jacobian determinants \cite[\S 16]{landau5}, we can express the heat capacity of the gas as an inner product:
\begin{equation}
c_G=\left(\dfrac{\partial \varepsilon}{\partial T}\right)_n=\dfrac{\partial(\varepsilon,n)}{\partial(T,n)}= \dfrac{\partial(\varepsilon,n)}{\partial(T,\alpha)} \dfrac{\partial(T,\alpha)}{\partial(T,n)}=\beta^3\left[(p^0\chi,p^0\chi)-\dfrac{(\chi,p^0\chi)^2}{(\chi,\chi)}\right]=\beta^3 (\varpi,\varpi) \, .
\end{equation}
Another useful quantity is the isothermal energy per added particle of the gas, namely
\begin{equation}
\curlyE=\left(\dfrac{\partial \varepsilon}{\partial n}\right)_T=\dfrac{(\partial\varepsilon/\partial \alpha)_T}{(\partial n/\partial \alpha)_T}= \dfrac{(\chi,p^0 \chi)}{(\chi,\chi)} \, .
\end{equation}

\section{Quadratic free-energy density in local equilibrium}\label{bbb}

In this manuscript, what we have been loosely referring to as the ``free-energy density'' $\mathfrak{F}$ is not actually the Helmholtz free-energy density. Rather, it is the density of the appropriate thermodynamic quantity that is minimized in equilibrium (sometimes, this quantity is called ``exergy'' \cite{BecattiniExergy2023}). In order to calculate it, one should follow the information-current procedure \cite{GavassinoGibbs2021,GavassinoCausality2021,GavassinoUniveraalityI2023odx}. Below, we show what this evaluates to for local-equilibrium perturbations.

For a system in contact with an external bath held at fixed temperature $T_\star$ and chemical potential $\mu_\star$, the appropriate density is
\begin{equation}
\mathfrak{F}=\varepsilon-T_\star s-\mu_\star n \, .
\end{equation}
Note that $\varepsilon$, $n$, and $s$ refer to the system of interest, while $T_\star$ and $\mu_\star$ refer to the bath. We now consider a one-parameter family of local-equilibrium states, parametrized by $\lambda$, while holding the parameters of the bath fixed. We take $\lambda=0$ to be the state of full equilibrium with the bath ($T=T_\star$, $\mu=\mu_\star$). Differentiating twice with respect to $\lambda$, we obtain (notation: $\dot{A}=dA/d\lambda$)
\begin{equation}
\begin{split}
\dot{\mathfrak{F}}={}& (T-T_\star)\dot{s}+(\mu-\mu_\star)\dot{n} \, , \\
\ddot{\mathfrak{F}}={}& (T-T_\star)\ddot{s}+(\mu-\mu_\star)\ddot{n}+\dot{T}\dot{s}+\dot{\mu}\dot{n} \, .
\end{split}
\end{equation}
Evaluating these expressions at $\lambda=0$, we obtain $\dot{\mathfrak{F}}=0$ and $\ddot{\mathfrak{F}}=\dot{T}\dot{s}+\dot{\mu}\dot{n}$. Thus, the quadratic perturbation $\Delta\mathfrak{F}=\frac{1}{2}\ddot{\mathfrak{F}}\lambda^2$ is related to the linear fluctuations $\delta A=\dot{A}\lambda$ of the thermodynamic quantities via
\begin{equation}
2\Delta\mathfrak{F}=\delta T\,\delta s+\delta\mu\,\delta n \, .
\end{equation}
Finally, expressing $\delta s$ and $\delta\mu$ in terms of $\delta T$ and $\delta n$,
\begin{equation}
\begin{split}
\delta s={}& \dfrac{c_V}{T}\delta T+\left(\dfrac{\partial s}{\partial n}\right)_T\delta n \, , \\
\delta\mu={}& \left(\dfrac{\partial\mu}{\partial T}\right)_n\delta T+\left(\dfrac{\partial\mu}{\partial n}\right)_T\delta n \, ,
\end{split}
\end{equation}
and using the Maxwell relation that follows from the differential of the Helmholtz free-energy density, $df_H{=}{-}s\,dT{+}\mu\,dn$,
\begin{equation}
\left(\dfrac{\partial\mu}{\partial T}\right)_n=-\left(\dfrac{\partial s}{\partial n}\right)_T \, ,
\end{equation}
we obtain
\begin{equation}
2\Delta\mathfrak{F}=\dfrac{c_V}{T}(\delta T)^2+\left(\dfrac{\partial\mu}{\partial n}\right)_T(\delta n)^2 \, ,
\end{equation}
as required.

\bibliography{Biblio}

\label{lastpage}
\end{document}